\documentclass[preprint,12pt]{elsarticle} 
\usepackage{graphicx}
\usepackage{amssymb}
\usepackage{amsmath}
\usepackage{amsfonts}
\usepackage{slashed}
\usepackage[dvipsnames]{xcolor}
\def\lamb#1#2{$^{#1}_{\Lambda}${#2}}
\def\lam#1#2{$^{#1}_{~\Lambda}${#2}}
\def\la#1#2{$^{#1}_{~~\Lambda}${#2}}

\newcommand{\be}{\begin{equation}} 
\newcommand{\ee}{\end{equation}}

\begin{document} 

\begin{frontmatter} 

\title{Constraints from $\Lambda$ hypernuclei on the $\Lambda NN$ content of 
the $\Lambda$-nucleus potential} 

\author{E.~Friedman} 
\author{~A.~Gal\corref{cor1}~}
\address{Racah Institute of Physics, The Hebrew University,
Jerusalem 91904, Israel} 
\cortext[cor1]{corresponding author: Avraham Gal, avragal@savion.huji.ac.il} 

\date{\today}

\begin{abstract}
A depth of $D_{\Lambda}\approx -28$ MeV for the $\Lambda$-nucleus potential 
was confirmed in 1988 by studying $\Lambda$ binding energies deduced from 
$(\pi^+,K^+)$ spectra measured across the periodic table. Modern two-body 
hyperon-nucleon interaction models require additional interaction terms, 
most likely $\Lambda NN$ three-body terms, to reproduce $D_{\Lambda}$. 
Here we apply a suitably constructed $\Lambda$-nucleus density dependent 
optical potential to binding energy calculations of observed $1s_{\Lambda}$ 
and $1p_{\Lambda}$ states in the mass range $12\leq A\leq 208$. The resulting 
$\Lambda NN$ contribution to $D_{\Lambda}$, about 14 MeV repulsion at 
symmetric nuclear matter density $\rho_0=0.17$~fm$^{-3}$, makes $D_{\Lambda}$ 
increasingly repulsive at $\rho\gtrsim 3\rho_0$, leading possibly to little 
or no $\Lambda$ hyperon content of neutron-star matter. This suggests in some 
models a stiff equation of state that may support two solar-mass neutron 
stars. 
\end{abstract}


\end{frontmatter}

\section{Introduction} 
\label{sec:intro} 

\begin{figure}[thb]
\begin{center} 
\includegraphics[width=0.96\textwidth]{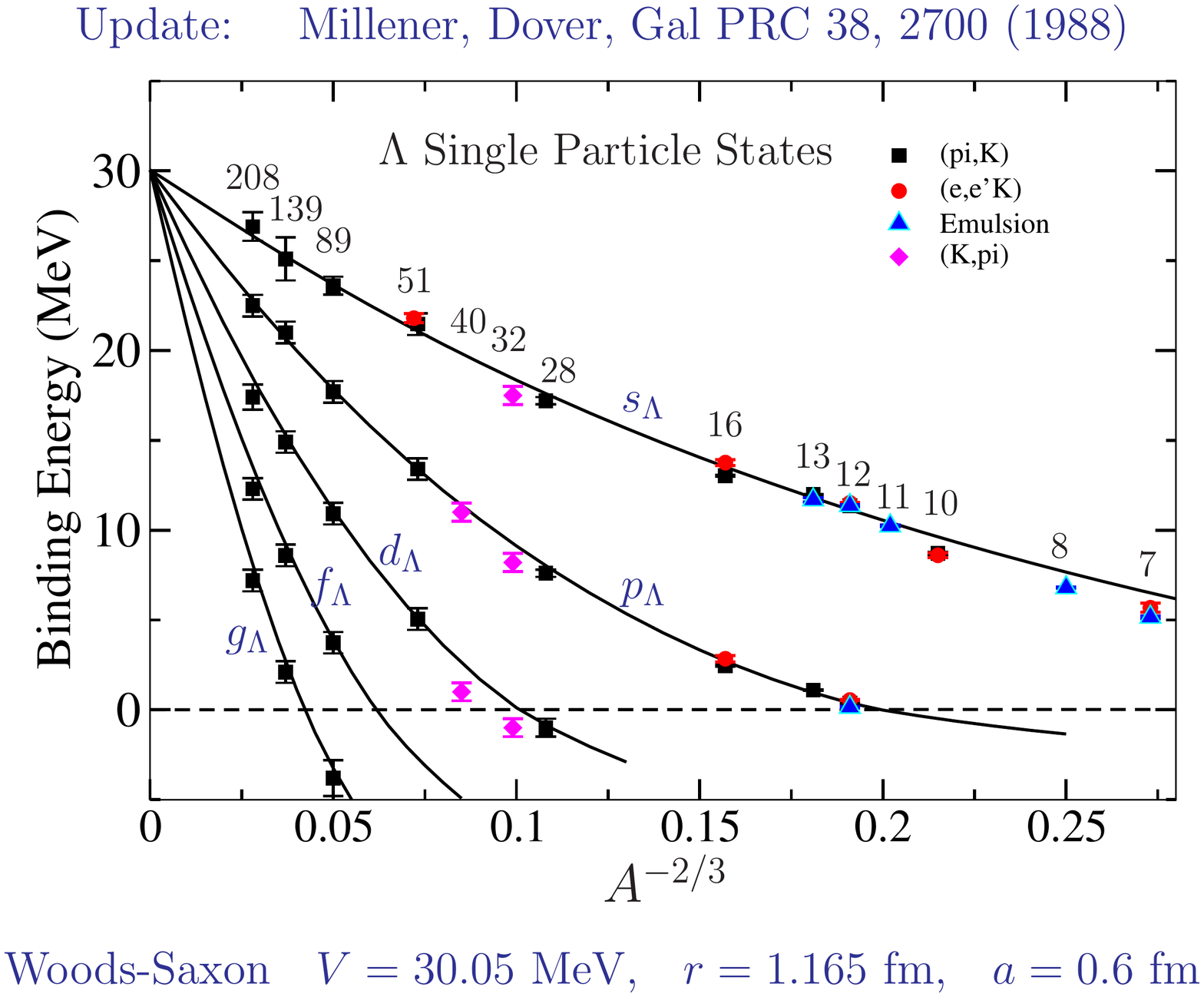} 
\caption{Compilation of $\Lambda$ binding energies in \lamb{7}{Li} to 
\la{208}{Pb} from various sources, and as calculated using a three-parameter 
WS potential \cite{MDG88}. Figure adapted from Ref. \cite{GHM16}.}
\label{fig:mdg}
\end{center} 
\end{figure}

The $\Lambda$-nucleus potential depth provides an 
important constraint in ongoing attempts to resolve the `hyperon puzzle', 
i.e., whether or not dense neutron-star matter contains hyperons, 
primarily $\Lambda$s besides nucleons \cite{FT20}. Fig.~\ref{fig:mdg} 
presents compilation of most of the known $\Lambda$ hypernuclear 
binding energies ($B_\Lambda$) across the periodic table, fitted by 
a three-parameter Woods-Saxon (WS) attractive potential. As $A\to\infty$, 
a limiting value of $B_{\Lambda}^{\rm exp}(A)\to 30$~MeV is obtained. 
This updates the value 28~MeV from the 1988 first theoretical analysis 
\cite{MDG88} of the BNL-AGS $(\pi^+,K^+)$ data \cite{Pile91}, and 27$\pm$3~MeV 
\cite{Lagnaux64} 27.2$\pm$1.3~MeV \cite{Lemonne65} from earlier mid 1960s 
observations of $\pi^-$ decays of heavy spallation hypernuclei formed in 
silver and bromine emulsions. Interestingly, studies of density dependent 
$\Lambda$-nuclear optical potentials $V_{\Lambda}(\rho)$ in Ref.~\cite{MDG88} 
conclude that a $\rho^2$ term motivated by three-body $\Lambda NN$ interactions 
provides a large repulsive (positive) contribution to the $\Lambda$-nuclear 
potential depth $D_{\Lambda}$ at nuclear-matter density $\rho_0$: 
$D_{\Lambda}^{(3)}\approx 30$~MeV. This repulsive component of 
$D_{\Lambda}$ is more than just compensated at $\rho_0$ by a roughly 
twice larger attractive depth value, $D_{\Lambda}^{(2)}\approx -60$~MeV, 
motivated by a two-body $\Lambda N$ interaction. 
Note that $D_{\Lambda}=D_{\Lambda}^{(2)}+D_{\Lambda}^{(3)}$ is defined as 
$V_{\Lambda}(\rho_0)$ in the limit $A\to\infty$ at a given nuclear-matter 
density $\rho_0$, with a value 0.17~fm$^{-3}$ assumed here. 

Most hyperon-nucleon potential models overbind $\Lambda$ hypernuclei, 
yielding values of $D_{\Lambda}^{(2)}$ deeper than $-30$~MeV. Whereas such 
overbinding amounts to only few MeV in the often used Nijmegen soft-core model 
versions NSC97e,f \cite{NSC97} it is considerably stronger, by more than 10 
MeV, in the recent Nijmegen extended soft-core model ESC16 \cite{ESC16}. 
A similar overbinding arises at leading order in chiral effective field theory 
($\chi$EFT) \cite{LO06}. The situation at next-to leading order (NLO) is less 
clear owing to a strong dependence of $D_{\Lambda}^{(2)}$ on the momentum 
cutoff scale $\lambda$ \cite{HV20}. At $\lambda$=500~MeV/c, however, it is 
found in Ref.~\cite{GKW20} that both versions NLO13 \cite{NLO13} and NLO19 
\cite{NLO19} overbind by a few MeV. Finally, recent Quantum Monte Carlo (QMC) 
calculations \cite{LGP13,LPG14}, using a $\Lambda N+ \Lambda NN$ interaction 
model designed to bind correctly \lamb{5}{He}, result in a strongly attractive 
$D_{\Lambda}^{(2)}$ of order $-100$~MeV and a correspondingly large repulsive 
(positive) $D_\Lambda^{(3)}$, reproducing the overall potential depth 
$D_{\Lambda}\approx -30$ MeV. 

Introducing repulsive three-body $\Lambda NN$ interactions goes beyond just 
providing solution of the overbinding problem: as nuclear density is increased 
beyond nuclear matter density $\rho_0$, the balance between attractive 
$D_\Lambda^{(2)}$ and repulsive $D_\Lambda^{(3)}$ tilts towards the latter. 
This results in nearly total expulsion of $\Lambda$ hyperons from neutron-star 
matter, suggesting an equation of state (EoS) sufficiently stiff to support 
two solar-mass neutron stars, thereby providing a possible solution to the 
`hyperon puzzle'. The larger $D_\Lambda^{(3)}$ is, the more likely it is a 
solution \cite{LLGP15,LVB19}. However, there is no guarantee that three-body 
$\Lambda NN$ interactions are universally repulsive. For a recent discussion 
of this problem within an SU(3) `decuplet dominance' approach practised in 
modern $\chi$EFT studies at NLO, see Ref.~\cite{GKW20}. 

Our aim in the present phenomenological study is to check to what extent 
properly chosen $\Lambda$ hypernuclear binding energy data, with minimal 
extra assumptions, imply positive values of $D_\Lambda^{(3)}$, and how 
large it is. To do so, we follow the optical potential approach applied 
by Dover-H\"{u}fner-Lemmer to pions in nuclear matter \cite{DHL71}. 
Applied to the $\Lambda$-nucleus system, it provides expansion in powers 
of the nuclear density $\rho(r)$, consisting of a linear term induced by 
a two-body $\Lambda N$ interaction plus two higher-power density terms: 
(i) a long-range Pauli correlations term starting at $\rho^{4/3}$, and 
(ii) a short-range $NN$ interaction term dominated in the present context 
by three-body $\Lambda NN$ interactions, starting at $\rho^2$. As will 
become clear below, the contribution of the Pauli correlations term is 
non negligible, propagating to higher powers of density terms than just 
$\rho^{4/3}$, such as the $\rho^2$ $\Lambda NN$ interaction term. This 
explains why the value derived here, $D_\Lambda^{(3)}=(13.9\pm 1.4)$~MeV, 
differs from any of those suggested earlier in Ref.~\cite{MDG88} 
and in Skyrme Hartree Fock studies \cite{SH14} where Pauli correlations 
apparently were disregarded. Our value of $D_\Lambda^{(3)}$ strongly disagrees 
with the much larger value inferred in QMC calculations \cite{LPG14}. 
We comment on this discrepancy below. 

The paper is organized as follows. In Sect.~2 we present the form of the 
optical potential $V_{\Lambda}^{\rm opt}(\rho)$ used to evaluate $B_{\Lambda}$ 
values across the periodic table. The choice of $B^{\rm exp}_{\Lambda}$ data 
and nuclear densities $\rho$, providing input to the present $B_{\Lambda}$ 
calculations, is briefly discussed. Results are given and discussed in 
Sect.~3, and concluding remarks are made in Sect.~4.

\section{Optical Potential Methodology and Input}
\label{sec:method}

The optical potential employed in this work, 
$V_{\Lambda}^{\rm opt}(\rho)=V_{\Lambda}^{(2)}(\rho)+V_{\Lambda}^{(3)}(\rho)$, 
consists of terms representing two-body $\Lambda N$ and three-body $\Lambda 
NN$ interactions, respectively: 
\begin{equation} 
V_{\Lambda}^{(2)}(\rho) = -\frac{4\pi}{2\mu_{\Lambda}}f_A\,
C_{\rm Pauli}(\rho)\,b_0\rho, 
\label{eq:V2} 
\end{equation}  
\begin{equation} 
V_{\Lambda}^{(3)}(\rho) = +\frac{4\pi}{2\mu_{\Lambda}}\,f_A\,B_0\,
\frac{\rho^2}{\rho_0}, 
\label{eq:V3} 
\end{equation} 
with $b_0$ and $B_0$ strength parameters in units of fm ($\hbar=c=1$). 
In these expressions, $A$ is the mass number of the nuclear core, $\rho$ 
is a nuclear density normalized to $A$, $\rho_0=0.17$~fm$^{-3}$ stands for 
nuclear matter density, $\mu_{\Lambda}$ is the $\Lambda$-nucleus reduced mass 
and $f_A$ is a kinematical factor transforming $b_0$ from the $\Lambda N$ c.m. 
system to the $\Lambda$-nucleus c.m. system: 
\begin{equation} 
f_A=1+\frac{A-1}{A}\frac{\mu_{\Lambda}}{m_N}. 
\label{eq:fA} 
\end{equation} 
This form of $f_A$ coincides with the way it is used for $V_{\Lambda}^{(2)}$ 
in atomic/nuclear hadron-nucleus bound-state problems~\cite{FG07} and its 
$A$ dependence provides good approximation for $V_{\Lambda}^{(3)}$. Next 
in Eq.~(\ref{eq:V2}) is the density dependent Pauli correlation function 
$C_{\rm Pauli}(\rho)$: 
\begin{equation} 
C_{\rm Pauli}(\rho)=(1+\alpha_P\frac{3k_F}{2\pi}f_Ab_0)^{-1}, 
\label{eq:Cpauli} 
\end{equation} 
with Fermi momentum $k_F=(3{\pi^2}\rho/2)^{1/3}$. The parameter $\alpha_P$ 
in Eq.~(\ref{eq:Cpauli}) switches off ($\alpha_P$=0) or on ($\alpha_P$=1) 
Pauli correlations in a form suggested in Ref.~\cite{WRW97} and practised in 
$K^-$ atoms studies \cite{FG17}. Following Ref.~\cite{FG21} we approximated 
$f_A$ in $C_{\rm Pauli}(\rho)$ by $f_{A\to\infty}$. As shown below, including 
$C_{\rm Pauli}(\rho)$ in $V_{\Lambda}^{(2)}$ affects strongly the balance 
between the derived potential depths $D_\Lambda^{(2)}$ and $D_\Lambda^{(3)}$. 
However, introducing it also in $V_{\Lambda}^{(3)}$ is found to make little 
difference, which is why it is skipped in Eq.~(\ref{eq:V3}). Finally we note 
that the low-density limit of $V_{\Lambda}^{\rm opt}$ requires according 
to Ref.~\cite{DHL71} that $b_0$ is identified with the c.m. $\Lambda N$ 
spin-averaged scattering length (positive here). We now specify the 
$B_{\Lambda}$ data dealt with in the present analysis and the nuclear 
densities $\rho(r)$ used for constructing the density dependent optical 
potential $V_{\Lambda}(\rho)$, Eqs.~(\ref{eq:V2},\ref{eq:V3}).

\subsection{$B_{\Lambda}$ Data}
\label{subsec:B_L}

The present work does not attempt to reproduce 
the full range of $B_{\Lambda}$ data shown in Fig.~\ref{fig:mdg}. 
It is limited to $1s_{\Lambda}$ and $1p_{\Lambda}$ states. We fit to 
such states in {\it one} of the nuclear $p$-shell hypernuclei where the 
$1s_{\Lambda}$ state is bound by over 10 MeV, while the $1p_{\Lambda}$ 
state has just become bound. This helps to resolve the density dependence 
of $V_{\Lambda}^{\rm opt}$ by setting a good balance between its two 
components, $V_{\Lambda}^{(2)}(\rho)$ and $V_{\Lambda}^{(3)}(\rho)$, and 
follow it throughout the periodic table up to the heaviest hypernucleus of 
\la{208}{Pb} marked in Fig.~\ref{fig:mdg}. Among the relevant $A=12,13,16$ 
$p$-shell hypernuclei, we chose to fit the \lam{16}{N} precise $B^{\rm exp}_{
\Lambda}(1s,1p)$ values, partly because of the extremely simple $1p$ proton 
hole structure of its nuclear core which removes most of the uncertainty 
arising from spin-dependent residual $\Lambda N$ interactions \cite{Mill08}. 

The $B^{\rm exp}_{\Lambda}(1s,1p)$ values considered in the present work are 
those shown in Fig.~\ref{fig:mdg} for $12\leq A\leq 208$, as listed in 
Table~IV of Ref.~\cite{GHM16} and remarked on in the related text. Most of 
these values are from $(\pi^+,K^+)$ reactions. Older $(K^-,\pi^-)$ data for 
\lam{32}{S} were included to enhance the only $(\pi^+,K^+)$ data available in 
the $2s$-$1d$ shell, for \lam{28}{Si}. For $A=12,16$ we used the more precise 
$B^{\rm exp}_{\Lambda}(1s,1p)$ values extracted in $(e,e'K^+)$ reactions at 
JLab \cite{JLab19}, \lam{12}{B} in preference to \lam{12}{C} and \lam{16}{N} 
instead of \lam{16}{O}. Whereas the \lam{12}{C} values agree with the 
\lam{12}{B} respective values within measurement uncertainties, this is not 
the case for the $A=16$ species where the $1s$ values differ by 0.8$\pm$0.3 
MeV. However, a more recent ($K^{-}_{\rm stop},\pi^-$) measurement on $^{16}$O 
reports $B_{\Lambda}^{1s}$(\lam{16}{N})=13.4$\pm$0.4 MeV \cite{Finuda11} 
consistently within its uncertainty range with the $(e,e'K^+)$ value 
$B_{\Lambda}^{1s}$(\lam{16}{N})=13.76$\pm$0.16 MeV \cite{JLab09} used here. 
We comment below on the sensitivity of our calculational results to this 
difference.

\subsection{Nuclear Densities} 
\label{subsec:rho} 

In optical model applications similar to the 
one adopted here, it is crucial to ensure that the radial extent of 
the densities, e.g., their r.m.s. radii follow closely values derived 
from experiment. With $\rho(r) = \rho_p(r) + \rho_n(r)$, the sum of 
proton and neutron density distributions, respectively, we relate 
the proton densities to the corresponding charge densities where 
the finite size of the proton charge and recoil effects are included. 
This approach is equivalent to assigning some finite range to the 
$\Lambda$-nucleon interaction. For the lightest elements in our database we 
used harmonic-oscillator type densities, assuming the same radial parameters 
also for the corresponding neutron densities \cite{Elton61}. For species 
beyond the nuclear $1p$ shell we used two-parameter and three-parameter 
Fermi distributions normalized to $Z$ for protons and $N=A-Z$ for neutrons, 
derived from nuclear charge distributions assembled in Ref.~\cite{AM13}.   
For medium-weight and heavy nuclei, the r.m.s. radii of neutron density 
distributions assume larger values than those for proton density 
distributions, as practiced in analyses of exotic atoms \cite{FG07}. 
Furthermore, once neutrons occupy single-nucleon orbits beyond those occupied 
by protons, it is useful to represent the nuclear density $\rho(r)$ as
\begin{equation} 
\rho(r)=\rho_{\rm core}(r)+\rho_{\rm excess}(r), 
\label{eq:exc1} 
\end{equation} 
where $\rho_{\rm core}$ refers to the $Z$ protons plus the charge symmetric 
$Z$ neutrons occupying the same nuclear `core' orbits, and $\rho_{\rm excess}$ 
refers to the $(N-Z)$ `excess' neutrons associated with the nuclear periphery.

\section{Results and Discussion} 
\label{sec:res} 

\begin{figure}[!h]
\begin{center} 
\includegraphics[height=12.0cm, width=0.96\textwidth]{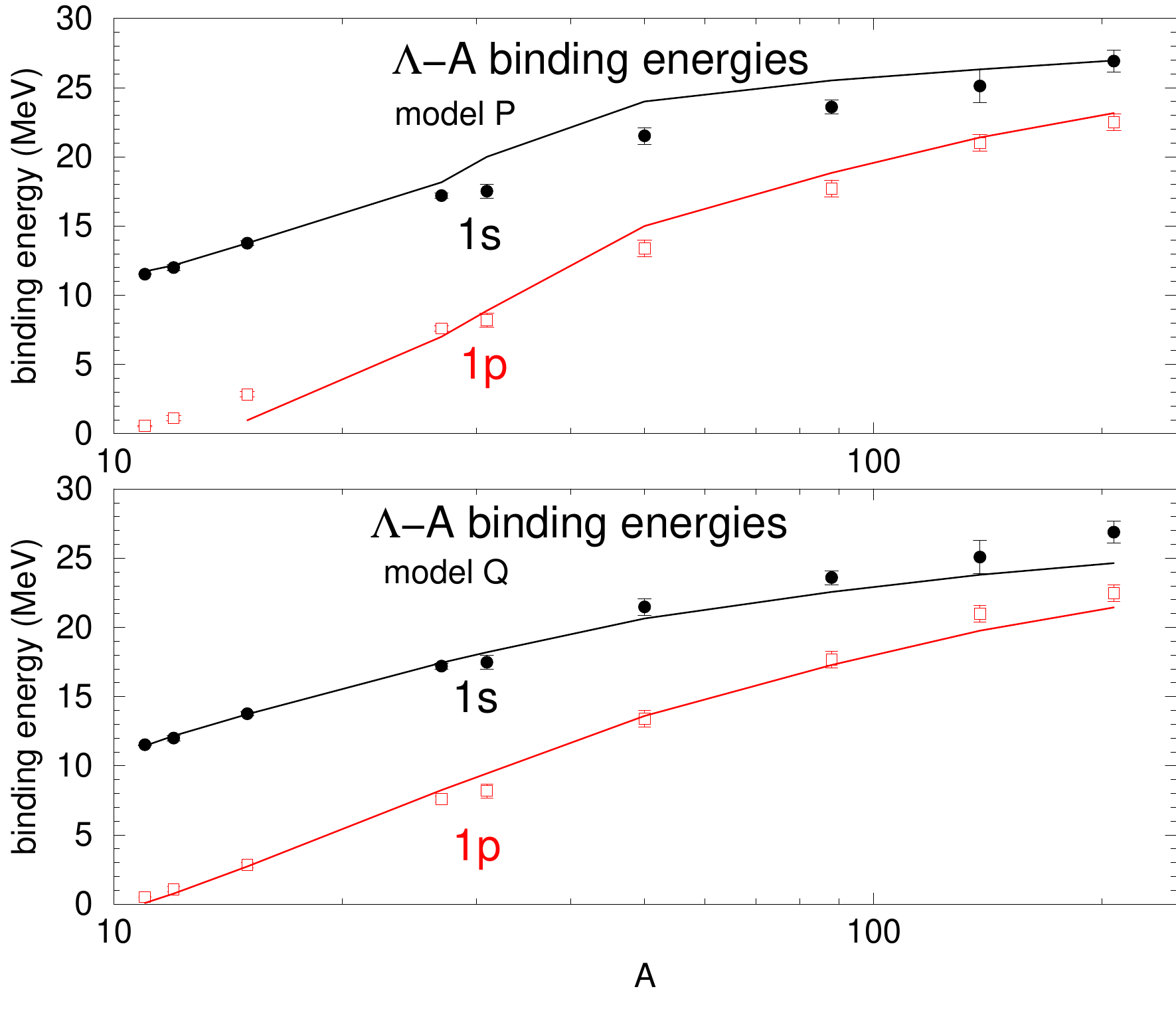} 
\caption{$B_{\Lambda}^{1s,1p}(A)$ values in Models P (upper) Q (lower) 
across the periodic table, see text. Data points, including uncertainties, 
are taken from Table IV of Ref.~\cite{GHM16}. Continuous lines connect 
calculated values.} 
\label{fig:modelsPQ} 
\end{center} 
\end{figure}

\begin{figure}[!t]
\begin{center} 
\includegraphics[height=12.0cm, width=0.96\textwidth]{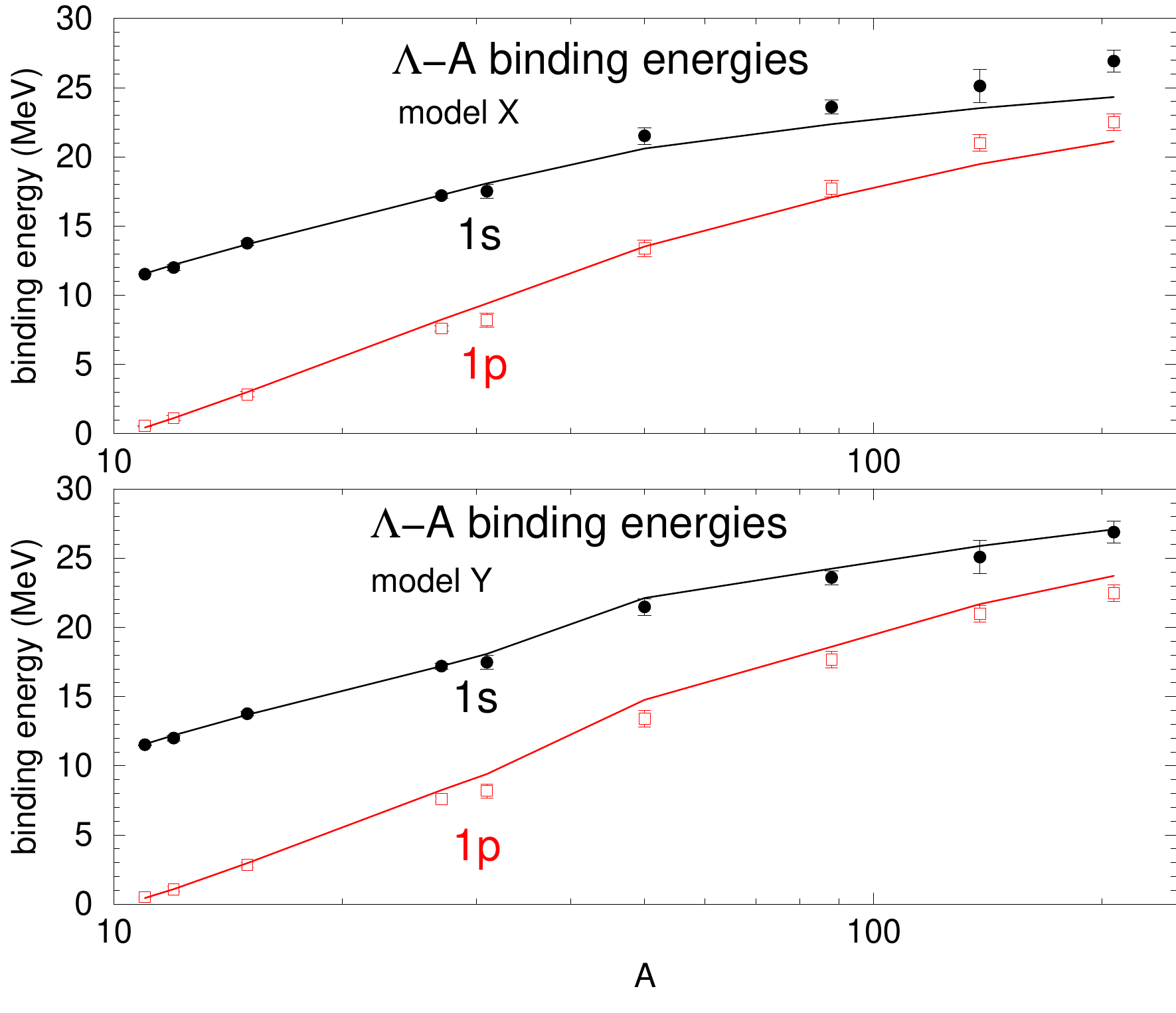} 
\caption{$B_{\Lambda}^{1s,1p}(A)$ values in Models X (upper) Y (lower) 
across the periodic table, see text. Data points, including uncertainties, 
are taken from Table IV of Ref.~\cite{GHM16}. Continuous lines connect 
calculated values.} 
\label{fig:modelsX2Y2}
\end{center} 
\end{figure}

$\Lambda$-nuclear potential depths 
$D_\Lambda^{(2)}$, $D_\Lambda^{(3)}$ from our optical-potential calculations, 
and their sum $D_\Lambda$, are listed in Table~\ref{tab:res} for several 
choices of $V_\Lambda^{\rm opt}$ models specified in detail below. Calculated 
$1s_{\Lambda}$ and $1p_{\Lambda}$ binding energies in hypernuclei from 
\lam{12}{B} to \la{208}{Pb} are shown in Fig.~\ref{fig:modelsPQ} for models 
P,Q, and in Fig.~\ref{fig:modelsX2Y2} for models X,Y, in comparison to 
$B^{\rm exp}_{\Lambda}(1s,1p)$ data. 

\begin{table}[!h] 
\begin{center} 
\caption{Strength parameters: (i) $b_0,B_0$ (fm), 
Eqs.~(\ref{eq:V2},\ref{eq:V3}), in models P,Q,X,Y of the present work, or (ii) 
$c_0,C_0$ (MeV$\cdot$fm$^3$, MeV$\cdot$fm$^6$) for $V_{\Lambda}(\rho)=-c_0\rho
+ C_0\rho^2$ from Ref.~\cite{MDG88}; plus their respective potential depths 
$D_{\Lambda}^{(2)}$, $D_{\Lambda}^{(3)}$ and sum $D_{\Lambda}$ (MeV) at 
nuclear matter density $\rho_0=0.17$~fm$^{-3}$. Pauli correlations are 
switched off (on) using $\alpha_P=0\,(1)$ in Eq.~(\ref{eq:Cpauli}).} 
\begin{tabular}{ccccccc} 
\hline 
Model & $\alpha_P$ & $b_0$ or $c_0$ & $B_0$ or $C_0$ & $D_{\Lambda}^{(2)}$ & 
$D_{\Lambda}^{(3)}$ & $D_{\Lambda}$ \\
\hline
P & 0 & 0.418 & -- & $-$34.1 & -- & $-$34.1 \\ 
P' & 1 & 0.908 & -- & $-$32.3 & -- & $-$32.3 \\ 
Q & 0 & 0.706 & 0.370 & $-$57.6 & 30.2 & $-$27.4 \\ 
MDG \cite{MDG88} & 0 & 340.0 & 1087.5 & $-$57.8 & 31.4 & $-$26.4 \\ 
X,Y & 1 & 1.85 & 0.170 & $-$41.6 & 13.9 & $-$27.7 \\  
\hline 
\end{tabular}
\label{tab:res} 
\end{center} 
\end{table} 

In Model P, the optical potential $V_{\Lambda}(\rho)$ consists of just 
a linear-density $V_{\Lambda}^{(2)}(\rho)$ term, disregarding Pauli 
correlations ($\alpha_P$=0). Its only strength parameter $b_0$ is fitted 
to $B_{\Lambda}^{1s}$(\lam{16}{N})=13.76$\pm$0.16 MeV. This and other 
$B_\Lambda(1s,1p)$ values calculated in Model P are plotted in the upper 
part of Fig.~\ref{fig:modelsPQ}. The model does well for $1s_{\Lambda}$ 
states below \lam{16}{N} and then in \la{208}{Pb}, but not in between. For 
$1p_{\Lambda}$ states it misses seriously the measured binding energies below 
\lam{28}{Si}, while doing fairly well in heavier species. The potential depth 
$D_{\Lambda}$ (here at $\rho_0=0.17$~fm$^{-3}$) associated with Model P is 
$-$34.1 MeV ($-$32.1 MeV at $\rho_0=0.16$~fm$^{-3}$), decreasing in size to 
$-$32.3 MeV upon switching on Pauli correlations ($\alpha_P$=1, Model P' in 
the table). The roughly 5\% decrease in size of $D_{\Lambda}$ is comparable 
to the 10\% decrease in size of $D_{\Xi}$ found recently for $\Xi^-$ 
hyperons~\cite{FG21}. 

In Model Q, a $V_{\Lambda}^{(3)}(\rho)$ $\rho^2$ component is added to 
the $V_{\Lambda}^{(2)}(\rho)$ linear-density component, but still with no 
Pauli correlations ($\alpha_P$=0). The two strength parameters $b_0,B_0$ are 
obtained by fitting to $B_{\Lambda}^{1s}$(\lam{16}{N})=13.76$\pm$0.16 MeV 
and $B_{\Lambda}^{1p}$(\lam{16}{N})=2.84$\pm$0.18 MeV. These and other $B_{
\Lambda}(1s,1p)$ values calculated in Model Q are plotted in the lower part 
of Fig.~\ref{fig:modelsPQ}. The overall fit quality is a bit improved with 
respect of that in Model P, but some underbinding appears to develop upon 
increasing the mass number $A$, noticed clearly in the three heaviest 
$1s_{\Lambda}$ and two heaviest $1p_{\Lambda}$ states. The resulting $\Lambda$ 
potential depth $D_{\Lambda}=-27.4$ MeV reflects a sizable cancellation 
between a strongly attractive two-body potential depth $D_{\Lambda}^{(2)}$ 
and a strongly repulsive three-body potential depth $D_{\Lambda}^{(3)}$ 
listed in Table~\ref{tab:res}. Interestingly, these two partial depths 
almost coincide with those listed in the following line of the table marked 
Model MDG after Ref.~\cite{MDG88}. 

Pauli correlations are switched on ($\alpha_P$=1) in Models X and Y. Model X 
uses the same densities $\rho(r)$ and $\rho^2(r)$ used in Models P,Q. We note 
in Table~\ref{tab:res} the steady increase of the fitted two-body strength 
parameter $b_0$ by roughly factor of two going from Model P to P', where Pauli 
correlations were switched on, or to Q where a three-body $\rho^2(r)$ term was 
introduced, and finally by roughly factor of four in Model X upon including 
both Pauli correlations and a three-body term. The fitted value in Model X, 
$b_0=1.85$~fm, is close to the spin averaged value of the $\Lambda N$ 
scattering length suggested by experiments (e.g., 1.65 fm \cite{Alex68} or 
1.78 fm \cite{HIRES10}). This indicates that once both Pauli correlations and 
a three-body term are taken into account, the present optical potential form 
satisfies approximately its $\rho\to 0$ constraint. We also note that the 
fitted three-body strength parameter $B_0$ is down by more than factor of two 
from its fitted value in Model Q where Pauli correlations were disregarded. 
It is a simple matter to check that the density expansion of $C_{\rm Pauli}
(\rho)$ produces repulsion at order $\rho^2$ which substitutes then for part 
of the value of $B_0$ in Model Q. Yet, Model X is unsatisfactory with respect 
to its calculated $B_{\Lambda}(1s,1p)$ values plotted in the upper part of 
Fig.~\ref{fig:modelsX2Y2}. The slight $1s_{\Lambda}$ underbinding observed in 
Model Q for $A\gtrsim 90$ has become more pronounced, beginning already with 
$A\gtrsim 50$, joined there now by considerable $1p_{\Lambda}$ underbinding. 
To address these shortcomings we introduce Model Y.  

Model Y differs from Model X by the form of $\rho^2$ used in nuclei where 
excess neutrons occupy shell-model orbits higher than those occupied by 
protons. This situation occurs in Fig.~\ref{fig:modelsX2Y2} for the four 
hypernuclei with $A\gtrsim 50$. 
We recall that the $\Lambda NN$ interaction underlying the $V_{\Lambda}^{(3)}$ 
$\rho^2$ term arises in $\chi$EFT models \cite{GKW20} from intermediate 
$\Sigma NN$ and $\Sigma^{\ast}(1385)NN$ states, yielding a ${\vec\tau}_1\cdot
{\vec\tau}_2$ isospin factor for the two nucleons \cite{Spitzer58,BLN67}. 
Extending the discussion in Ref.~\cite{GSD71}, it can be shown that direct 
three-body $\Lambda NN$ contributions involving one `core' nucleon and 
one `excess' nucleon vanish upon summing on the $T$=0 `core' closed-shell 
nucleons, while their exchange partners renormalize the two-body $\Lambda N$ 
interaction. To modify $\rho^2$ accordingly, we discard the bilinear 
term $\rho_{\rm core}\rho_{\rm excess}$, thereby replacing $\rho^2$ in 
$V^{(3)}_{\Lambda}$, Eq.~(\ref{eq:V3}), by 
\begin{equation} 
\rho_{\rm core}^2+\rho_{\rm excess}^2\rightarrow(2\rho_p)^2+(\rho_n-\rho_p)^2 
\label{eq:exc2} 
\end{equation}
in terms of the available densities $\rho_p$ and $\rho_n$. 
This prescription does not impact the five hypernuclei lighter than $^{40}$Ca 
in Fig.~\ref{fig:modelsX2Y2}, and it appears to work well as seen by the 
significant improvement in the $B_{\Lambda}(1s)$ values calculated 
beyond $^{40}$Ca in Model Y, lower part of Fig.~\ref{fig:modelsX2Y2}, 
compared to those in Model X in the upper part of the figure. 

As a simple check on the Model Y results exhibited in 
Fig.~\ref{fig:modelsX2Y2}, we apply a scaling-factor 
fraction $F$ to the $\rho^2$ term of Model X, 
\begin{equation} 
F=\frac{(2Z)^2+(N-Z)^2}{A^2}.  
\label{eq:F} 
\end{equation} 
This fraction is the reduced $\rho^2$ volume integral of a core of $Z$ protons 
and their charge symmetric $Z$ neutrons, and excess of ($N-Z$) neutrons, out 
of the $A$ nucleons. The resulting $B_{\Lambda}$ values in this simplified 
model indeed agree closely with those in Model Y shown in the lower part of 
the figure. Interestingly, if this scaling factor $F$ is applied in Model Q 
where $D^{(3)}_{\Lambda}$ is more than twice larger than in Model Y (see 
Table~\ref{tab:res}), $\Lambda$ single-particle states in the four heaviest 
hypernuclei shown in the lower panel of Fig.~\ref{fig:modelsPQ} become 
substantially overbound, e.g. $B_{\Lambda}^{1s}$(\la{208}{Pb})=32.0~MeV. 
This demonstrates the importance of including Pauli correlations in the 
$V_{\Lambda}^{(2)}$ component of the $\Lambda$-nucleus optical potential. 

Potential depth values derived in models discussed above are listed in 
Table~\ref{tab:res}. Model Y in particular gives $D^{(2)}_{\Lambda}=-41.6
$~MeV, $D^{(3)}_{\Lambda}=13.9$~MeV. To estimate uncertainties, we act as 
follows: (i) decreasing the input value of $B_{\Lambda}^{1s}$(\lam{16}{N}) 
fitted to by 0.2~MeV, thereby getting halfway to the central value of $B_{
\Lambda}^{1s}$(\lam{16}{O})=13.4$\pm$0.4~MeV for \lam{16}{O} \cite{Finuda11} 
the charge symmetric partner of \lam{16}{N}, results in approximately 10\% 
larger value of $D^{(3)}_{\Lambda}$, and (ii) applying Pauli correlations 
to $V^{(3)}_{\Lambda}$ too reduces $D^{(3)}_{\Lambda}$ roughly by 10\%. 
In both cases $D^{(2)}_{\Lambda}$ increases moderately by $\lesssim$1~MeV. 
On the other hand, $D^{(2)}_{\Lambda}$ decreases by 1.7~MeV once the $A$ 
dependence of $f_A$, neglected in our present use of $C_{\rm Pauli}(\rho)$ 
Eq.~(\ref{eq:Cpauli}), is restored. Considering these uncertainties, 
our final values are (in MeV)  
\begin{equation} 
D_\Lambda^{(2)}=-(40.6\pm 1.0),\,\,\,\,\,\,D_\Lambda^{(3)}=(13.9\pm 1.4), 
\label{eq:D} 
\end{equation} 
and $D_{\Lambda}=-26.7\pm 1.7$~MeV. 

The $D^{(2)}_{\Lambda}$ and $D^{(3)}_{\Lambda}$ values in Eq.~(\ref{eq:D}) are 
considerably lower than those deduced in QMC calculations~\cite{LGP13,LPG14}. 
Note that the QMC nuclear densities $\rho_{\rm QMC}(r)$ are much too 
compact with respect to our realistic densities, with nucleon r.m.s. radii 
$r_N$(QMC) only $\sim$80\% of the known r.m.s. charge radii in $^{16}$O and 
$^{40}$Ca~\cite{Lon13}. Since $\rho$ scales as $r_N^{-3}$, applying it to the 
density dependence of our $V^{\rm opt}_{\Lambda}$ would transform $D^{(2)}_{
\Lambda}$ and $D^{(3)}_{\Lambda}$ of Eq.~(\ref{eq:D}) to $D^{(2)}_{\Lambda}
$(QMC)=$-$79.3$\pm$2.0 MeV and $D^{(3)}_{\Lambda}$(QMC)=53.0$\pm$5.3 MeV, 
their sum $D_{\Lambda}$(QMC)=$-$26.3$\pm$5.7 MeV agreeing within uncertainties 
with ours. Other factors such as the choice of fitted $B_\Lambda$ values may 
contribute to increase further the size of these QMC partial depth values to 
come closer to as large sizes as of ${\cal O}$(100)~MeV reached in these works.

\section{Concluding Remarks} 
\label{sec:concl} 

In summary, we have presented a straightforward 
optical-potential analysis of $1s_\Lambda$ and $1p_\Lambda$ binding energies 
across the periodic table, $12\leq A\leq 208$, based on nuclear densities 
constrained by charge r.m.s. radii. The potential is parameterized by 
constants $b_0$ and $B_0$ in front of two-body $\Lambda N$ and three-body 
$\Lambda NN$ interaction terms. These parameters were fitted to precise 
$B^{\rm exp}_{\Lambda}(1s,1p)$ values in \lam{16}{N} and then used to evaluate 
$B_{\Lambda}(1s,1p)$ values in the other hypernuclei considered here. Pauli 
correlations were found essential to establish a correct balance between $b_0$ 
and $B_0$, as judged by $b_0$ coming out in the final Model Y analysis close 
to the value of the $\Lambda N$ spin-averaged $s$-wave scattering length. 
Good agreement was reached in this model between the calculated 
$B^{1s,1p}_{\Lambda}$ values and their corresponding $B^{\rm exp}_{\Lambda}$ 
values; see Fig.~\ref{fig:modelsX2Y2}. Although values of $\ell_{\Lambda}$ 
other than $1s_{\Lambda}$ and $1p_{\Lambda}$ were disregarded, we checked that 
$B_{\Lambda}^{1d,1f,1g}$(\la{208}{Pb}) values calculated in Model Y come out 
reasonably well within 1-2 error bars of the experimental values shown in 
Fig.~\ref{fig:mdg}. 

The potential depth $D^{(3)}_{\Lambda}$ derived here, Eq.~(\ref{eq:D}), 
suggests that in symmetric nuclear matter the $\Lambda$-nucleus potential 
becomes repulsive near three times nuclear-matter density $\rho_0$. 
Our derived depth $D^{(3)}_{\Lambda}$ is larger by a few MeV than the 
one yielding $\mu(\Lambda) > \mu(n)$ for $\Lambda$ and neutron chemical 
potentials in purely neutron matter under a `decuplet dominance' construction 
for the underlying $\Lambda NN$ interaction terms within a $\chi$EFT(NLO) 
model~\cite{GKW20}. This suggests that the strength of the corresponding 
repulsive $V_\Lambda^{(3)}$ optical potential component, as constrained in 
the present work by data, is sufficient to prevent $\Lambda$ hyperons from 
playing active role in neutron-star matter, thereby enabling a stiff EoS 
that supports two solar-mass neutron stars.

\section*{Acknowledgments} 

We gratefully acknowledge useful remarks by J.~Mare\v{s}, D.J.~Millener, 
H.~Tamura, I.~Vida\~{n}a and W.~Weise. A preliminary report of the 
present work was presented at the HYP2022 International Conference in 
Prague~\cite{FG22} as part of a project funded by the European Union's 
Horizon 2020 research \& innovation programme, grant agreement 824093.

\end{document}